\def\BibTeX{{\rm B\kern-.05em{\sc i\kern-.025em b}\kern-.08em
            T\kern-.1667em\lower.7ex\hbox{E}\kern-.125emX}}

\def\\{\char'134}

\documentclass[letter,11pt]{article} 
\pdfoutput=1
\usepackage[margin=2cm]{geometry}
\usepackage[sumlimits,intlimits]{amsmath}
\usepackage{amsfonts,amssymb}
\usepackage{mathrsfs}
\usepackage{amsmath}
\usepackage[T1]{fontenc}
\usepackage{capt-of}
\usepackage{ucs}
\usepackage{graphics}
\usepackage[dvips]{graphicx}
\usepackage{textcomp}
\usepackage{amssymb}
\DeclareGraphicsExtensions{.pdf,.png,.eps}
\usepackage{lineno}
\usepackage[square,sort&compress,comma]{natbib}
\setcitestyle{numbers}

\usepackage{lineno}

\def\diagram#1{{\normallineskip=8pt\normalbaselineskip=0pt \begin{matrix}#1\end{matrix}}}

\def\harr#1#2{\smash{\mathop{\hbox to .3in{\rightarrowfill}}
 \limits^{\scriptstyle#1}_{\scriptstyle#2}}}

\def\ov{\overline}

\def\wh{\widehat}
\def\H{\text{H}}
\def\mb{\mathbf}
\def\wt{\widetilde}
\def\beq{\begin{equation}}
\def\eeq{\end{equation}}
\def\K{\mathscr{K}}
\def\W{\mathscr{W}}
\def\C{\wt{\cal C}_3}
\def\P{\wh\Pi_3}
\def\S{\Sigma_3^{tor}}
\def\a{\wh\alpha_0}
\def\b{\beta^{0,tor}}
\def\G{\wt\Gamma_3}
\def\e{\left(\frac{e^0}{k}~mod~1\right)}

\begin{document}
\makeatletter
\@addtoreset{equation}{section}
\makeatother
\renewcommand{\theequation}{\thesection.\arabic{equation}}


\vspace{.5cm}
\begin{center}
\Large{\bf  Half-flat  Quantum Hair}\\

\vspace{1cm}

\large
Hugo  Garc\'ia-Compe\'an$^{a, } $\footnote{e-mail:{\tt compean@fis.cinvestav.mx}},  Oscar Loaiza-Brito$^{b,} $\footnote{e-mail :{\tt oloaiza@fisica.ugto.mx}}, Aldo Mart\'inez-Merino$^{b,} $\footnote{e-mail :{\tt a.merino@fisica.ugto.mx}} and  Roberto Santos-Silva$^{b,} $\footnote{e-mail :{\tt rsantos@fisica.ugto.mx}} \\[4mm]

{\small \em $^a$Departamento de F\'isica, Centro de Investigaci\'on y de Estudios Avanzados del I.P.N.}\\
{\small\em P.O. Box 14-740, 07000, Mexico D.F., Mexico.}\\

\bigskip

{\small\em $^b$Departamento de F\'isica, Universidad de Guanajuato}\\
{\small\em  C.P. 37150, Le\'on,  Guanajuato, Mexico.}\\[4mm]
\vspace*{2cm}
\small{\bf Abstract} \\
\end{center}

\begin{center} 
\begin{minipage}[h]{14.0cm} {By wrapping $D3$-branes over 3-cycles on a half-flat manifold we construct an effective supersymmetric black hole in the ${\cal N}=2$ low-energy theory in four-dimensions. Specifically we find that the torsion cycles present in a half-flat compactification, corresponding to the mirror symmetric image of electric NS flux on a Calabi-Yau manifold, manifest in the half-flat black hole  as {\it quantum hair}. We compute the electric and magnetic charges related to the quantum hair, and also the mass contribution to the effective black hole. We find that by wrapping a number of $D3$-branes equal to the order of the discrete group associated to the torsional part of the half-flat homology, the effective charge and mass terms vanishes. We compute the variation of entropy  and the corresponding temperature associated with the lost of  quantum hair. We also comment on the equivalence between canceling Freed-Witten anomaly and the assumption of self-duality  for the 5-form field strength. Finally  from a K-theoretical perspective, we compute the presence of discrete RR charge of $D$-branes wrapping torsional cycles in a half-flat manifold.\\
}

\end{minipage} 
\end{center}

\newpage


\section{Introduction}

Inclusion of flux backgrounds in the search of string compactification scenarios   leading to fully symmetric four-dimensional spaces with some or all moduli stabilized, has been studied in a huge effort and detail in the last decade (see for example \cite{Grana:2005jc} and references therein).
Within this context, there has been enormous advances in the construction of string models in which (supersymmetric) Standard-Model like scenarios are immersed (\cite{Uranga:2007zza, Marchesano:2007de, RamosSanchez:2008tn}). \\

On the other hand, Supersymmetric Black Holes (SBH's) have been  constructed in the context of Type II
string theory fluxless compactifications on a Calabi-Yau manifolds by
wrapping $Dp$-branes on internal $p$-cycles \cite{Strominger:1995cz, Suzuki:1995rt, Strominger:1996kf, Shmakova:1996nz, Lust:1997kx, Bertolini:1998se, Denef:1999th, Denef:2000nb, Mohaupt:2000mj, Mohaupt:2000gc, Ooguri:2005vr, Pioline:2008zz} and in the corresponding low energy limit given by supergravity ${\cal N}=2$ \cite{Ferrara:1995ih, Behrndt:1996jn, Sabra:1997kq, Sabra:1997dh, Dall'Agata:2011nh, Hristov:2012bk}.  Integer electric and
magnetic charges are computed from the corresponding massless RR
potential fields.  However, as it is well known, string compactifications on Calabi-Yau manifolds are far from being realistic for various reasons among which  the moduli stabilization problem and the  {\it a priori} selection of vibrational modes of the quantum string.\\

Hence, it would be desirable to construct black holes on the grounds of a more general theory where (Ramond-Ramond) RR and (Neveu-Schwarz-Neveu-Schwarz) NS-NS fluxes are turned on. The main problem arises from the fact that in the most general background, since the fluxes are translated into the mirror symmetric manifold as torsion contributions to the Levi-Civita connection, most of the forms (as the holomorphic (3,0) form $\Omega$ and the K\"ahler form $J$) do not close under the standard differential operator $d$, and the relation among harmonic fields and standard cohomology is lost. Without this identification, effectie theories in 4d can not being constructed as easily as in the CY case.\\

Huge efforts have been made in the past years to construct ${\cal N}=2$ gauged supergravities from compactifications of Type II strings on generalized manifolds, as half-flat manifolds, such as internal spaces with $SU(3)$ or $SU(3)\times SU(3)$ structures, etc. Construction of black hole solutions within the context of gauged supergravity has also been studied in the last years \cite{Hristov:2009uj, Hristov:2010eu, Hristov:2011ye, Hristov:2011qr}. On the other hand, non-perturbative corrections to the prepotential in Type IIA string theory compactifications on Calabi-Yau manifolds and for self-mirror manifolds have been considered recently to find analytical  black hole solutions  involving non-extremal solutions as well as an interesting solution describing a supersymmetric black hole \cite{Bueno:2012jc, Bueno:2013psa}\\

It is important to remark that the inclusion of NS-NS fluxes in the construction of SBH could made some of the involved branes unstable to decay into closed strings as shown in \cite{LoaizaBrito:2007kz} due to the presence of a Freed-Witten (FW) anomaly \cite{Freed:1999vc, Maldacena:2001xj, Evslin:2006cj}. If the FW anomaly is cancelled, SBH constructed in such scenarios seems to be stable \cite{Danielsson:2006jg}. Nevertheless, the amount of fluxes considered  must be small otherwise we cannot assure a small back-reaction due to fluxes in the supergravity approach. We are interested precisely on this point and study the effects on the black hole by wrapping $D3$-branes on an internal manifold in which the back-reaction has been considered.\\

Therefore, one can foresee two options for constructing a SBH:  by wrapping $D3$-branes on a CY manifold threaded with a slight amount of NS-NS  flux, or by wrapping $D3$-branes on a generalized manifold  \cite{Hsu:2006vw, Danielsson:2006jg, LoaizaBrito:2007kz, Larfors:2009zz, LoaizaBrito:2010uw}.  The main goal of this work is to establish the first steps to study the physics of a black hole constructed by wrapping  $D$-branes on a generalized manifold.  .\\

Concretely we focus our study on the construction of SBH
by {\it wrapping $D$-branes on torsional cycles of a half-flat manifold} without considering quantum corrections to the superpotential or prepotential. In this sense, we are studying at tree level, how to construct SBH on manifolds which already have back-reacted to the presence of NS-NS fluxes. The back-reaction is then manifested  by the appearance of torsional components on which wrapped $D$-branes contribute to extra degrees of freedom to the low energy theory  identified with (a supersymmetric version of)  {\it quantum hair}
studied in \cite{ Preskill:1990bm, Preskill:1990ty, Coleman:1991ku, Coleman:1991sj, Banks:2010zn, Camara:2011jg, Veneziano:2012yj}, and measure it by  the presence of
a 4d string.\\

Our work is organized as follows: In section 2, we briefly review the standard construction of  a SBH in a Type IIB string CY compactification by wrapping $D3$-branes on supersymmetric 3-cycles. In Section 3, we compute and review some of the most important properties of a half-flat manifold including the derivation of  cohomology groups and the expansion of some fields in terms of torsional forms,  followed by the calculation of the electric and magnetic discrete charges and the torsional mass contribution. With this, we can compute the entropy variation of the state by increasing the number of torsional $D3$-branes and the associated temperature once the black hole losses its hair as the number of torsional branes reaches the order of the discrete group. Finally, in Section 4 we use K-theory in order to compute the discrete charge associated to D-branes in torsional cycles with the purpose to elucidate the nature of discrete charge without using an extra object as a 4d-string manifested as the Aharanov-Bohm effect. Our final comments are given in section 5 followed by a couple of Appendices. In Appendix A we refer to the usual notation for the symplectic cohomology basis, while in Appendix B we explicitly show a review on the construction of the low-energy  theory corresponding to a Type IIB compactification on a half-flat manifold on which the 5-form field strength is expanded in terms of torsional forms.

\section{Supersymmetric Black Holes from  wrapped $D3$-branes}
It is well known from the past years, that 
a SBH in four dimensions can be constructed by wrapping
D-branes in internal  non-trivial cycles. The physics of the effective BPS object can be derived from different approaches \cite{Strominger:1995cz, Suzuki:1995rt, Shmakova:1996nz,Bertolini:1998se,  Denef:1999th, Ooguri:2005vr, Larfors:2009zz}. According to our purposes, we would like to review the construction of a SBH in the Type IIB scenario in which D3-branes wrap internal 3-cycles of a CY manifold $X_3$, closely following \cite{Suzuki:1995rt}. \\

A massive SBH is obtained by wrapping a large number of D3-branes on
the corresponding cycles (otherwise they simply describe elementary
massive particles). The gauge field $\mathscr{A}_1$ related to the electric and magnetic charges in
the effective 4d ${\cal N}=2$ theory is constructed from  the self-dual RR field strength $\mathscr{F}_5$, given by  $\mathscr{F}_{5}=\mathscr{F}_2\wedge F_3$, with $d\mathscr{A}_1=\mathscr{F}_2$, and through the decomposition  driven by compactifying the extra 6-dimensions on $X_3$. \\

To see that, consider $N$ $D3$-branes wrapping an internal 3-cycle  ${\cal C}_3 \subset X_3$ given by  a linear combination of the symplectic basis of 3-cycles  $(A^I, B_I)$ with ${\cal PD}_6(A^I)=\beta^I$, ${\cal PD}_6(B_I)=\alpha_I$, and $I=0,1,\cdots ,h^{(2,1)}(X_3)$. The basis of  3-forms $(\alpha_I, \beta^I)$ is chosen to satisfy 
as usual
\begin{equation}
\int_{A^J}\alpha_I=-\int_{B_I}\beta^J=\delta^J_I.
\end{equation}
Defining the RR potential related to these $D3$-branes by
\beq
C_4=A_1\wedge \sum_I (e^I\alpha_I-m_I\beta^I),
\eeq
the (non self-dual) electric part of $\mathscr{F}_5$ can be written as
\begin{equation}
F_5= F_2\wedge F_3=F_2\wedge \sum_I(e^I\alpha_I -m_I\beta^I).
\end{equation}
The electric charge $Q_e$ is computed by integrating $*_{10}F_5$ over a 5-cycle $\Gamma_5$ identified as the boundary of $\Gamma_6=\mathbf{B}^3\times \Gamma_3$.  Actually, since $\Gamma_3$ belongs to $\text{H}_3(X_3;\mathbf{Z})$, only  the four-dimensional component of this cycle has boundary. Both, electric and magnetic charges are then given by
\begin{eqnarray}
Q_e&=&\int_{\mathbf{S}^2\times \Gamma_3}\star F_2\wedge {\cal PD}_6({\cal C}_3)=-qN,\nonumber\\
Q_m&=&\int_{\mathbf{S}^2\times {\cal C}_3}F_2\wedge {\cal PD}_6(\Gamma_3)=pN,
\end{eqnarray}
where ${\cal PD}_6({\cal C}_3)=\ast F_3$, ${\cal PD}_6(\Gamma_3)=F_3$ with the intersection number $\Gamma_3\cap {\cal C}_3=N$. From this it follows that
\beq
{\cal C}_3= -p_IA^I+q^IB_I,
\eeq
with
\begin{eqnarray}
q^I&=&e^JA_{J}^I-m_JC^{IJ},\nonumber\\
p_I&=&-e^JB_{IJ}-m_JA_I^J.
\end{eqnarray}
with the matrices $A, B$ and $C$ defined through integration of the wedge product between $(\alpha_I,\beta^I)$ and their duals as depicted in relations (\ref{matricesap}). The total charge of the system can also be computed by integrating the self-dual 5-form $\mathscr{F}_5=F^I\wedge\alpha_I-G_I\wedge\beta^I$ over  the cycle ${\cal C}_3\cup\Gamma_3$ as
\beq
Q_T=\int_{\mathbf{S}^2\times({\cal C}_3\cup\Gamma_3)}\mathscr{F}_5=N(p-q),
\eeq
where
\beq
F^I= e^IF_2+q^I\star F_2, \quad \text{and}\quad G_I=m_IF_2+p_I\star F_2.\\
\eeq

On the other hand, since $D3$-branes are BPS states of the theory, it is expected that the point-like object in the effective theory should be a BPS object as well. This means that it represents a massive state in the short-multiplet of the ${\cal N}=2$ supersymmetric theory with a metric given by \cite{Ferrara:1995ih}
\begin{equation}
ds^2=-e^{2U(\tau)}dt^2+\frac{e^{-2U(\tau)}}{\tau^4}d\tau^2+\frac{e^{-2U(\tau)}}{\tau^2}d\Omega^2,
\end{equation}
where $U(\tau)$ vainishes as $\tau\rightarrow 0$ and diverges at the horizon. Hence, the RR 5-form is self-dual in this metric provided
\begin{equation}
F_2=sin~\theta~d\theta\wedge ~d\phi \qquad \text{and} \qquad \star F_2=e^{2U}dt\wedge d\tau.
\end{equation}
The effective scalar potential $\mathscr{V}(r)$, computed by dimensionally reducing the ten-dimensional term $F_5\wedge *_{10}F_5$ (with the corresponding self-duality being imposed afterwards) reads
\beq
\mathscr{V}(r)=\tau^4\mathscr{V}_{BH},
\eeq
where,
\begin{eqnarray}
\mathscr{V}_{BH}&=&\int_{X_6}F_3\wedge\ast F_3=e^{\mathscr{K}}\left(D_I\mathscr{W}D_{\bar{J}}\bar{\mathscr{W}}K^{I\bar{J}}+3|\mathscr{W}|^2\right),\nonumber\\
&=&-e^Ip_I+m_Iq^I=N,
\label{VBH}
\end{eqnarray}
with the superpotential $\mathscr{W}$ given by 
\begin{equation}
\mathscr{W}=\int_{X_6} F_3\wedge \Omega_3.
\label{W}
\end{equation}
Being a BPS state, a SBH is extremal by construction.
The sum of the squared charges, $Q^2=Q_e^2+Q_m^2$ equals the mass of the expected supersymmetric BPS object in the four-dimensional ${\cal N}=2$ supergravity theory on which one obtains that $F_3=Re(C\Omega_3)$ for an arbitrary complex constant $C$  with $\Omega$ being the unique holomorphic $(3,0)$ form in $\text{H}^3(X_3\;\mathbf{Z})$. Although the SBH's mass is formally computed through the use of the special symplectic geometry \cite{Strominger:1990pd, Ferrara:1995ih, Ceresole:1995ca}, we shall reconstruct  it following the prescription given in \cite{Becker:1995kb, Denef:1999th} which fits our purposes better.\\

The mass can be directly computed from the Dirac-Born-Infeld action of the D3-branes wrapping ${\cal C}_3$,
\begin{equation}
S_{D3}=\int_{\gamma\times {\cal C}_3}\sqrt{-G}=-M_{BPS}\int_\gamma ds,
\end{equation}
where $G_{MN}$ is the world-volume metric of the $D3$-branes,  decomposing as $G_{MN}=\mathbf{1}\otimes g_{mn}$. After assuming preservation of supersymmetry in four-dimensions (which implies that ${\cal C}_3$ is a special Lagrangian cycle) it is possible to show that 
\begin{equation}
M^2_{BPS}=e^{\mathscr{K}}|\mathscr{W}|^2=\frac{1}{2Im(\bar{\tau}_{IJ}X^I\bar{X}^J)}\left|e_IX^I-m^IF_I\right|^2,
\end{equation}
where $\mathscr{K}=-\ln~ i\int \Omega_3\wedge \bar\Omega_3=-\ln~i(\bar{X}^IF_I-X^I\bar{F}_I)$ and $F_I=\tau_{IJ}X^J$.
As noticed in \cite{Suzuki:1995rt}, the total charge and the  mass are equal, as corresponding to a BPS object, by considering only the graviphoton mass. At the end of the day, we have a BPS point-like object with a horizon, within the extremal condition on which its charge equals its mass. It is interesting to notice that the black hole charge in four dimensions can be understood as a linking number among the 3-dimensional ball with $S^2$ as its boundary, and a point-like object. From the internal space point of view, the quantity $\int_{{\cal C}_3} F_3$ also represent a linking number\footnote{This is also reflected in the definition of Poincar\'e duals between $(A^I,B_I)$ and $(\alpha_I,\beta^I)$.} among the internal components of the RR 3-form and the cycle on which the integration is performed.  One could say after such observation that electric and magnetic charges of four-dimensional objects constructed from extended branes in higher dimensional spaces, correspond to an arrangement of those branes such that there is a linking number in 4 dimensions and in the internal space. However, the fact  the mass equal its charge is not so evident from this perspective, since the mass is computed through and integral which does not represent a linking number. Mass and charge are equal due to the fact that the cycles over which they are computed are supersymmetric. \\

Finally, the field content in the background theory with ${\cal N}=2$ in four dimensions is constructed from  an expansion of the 10d massless RR fields on a basis of  cohomological forms in the internal space in order to
describe massless states in four dimensions. The existence of scalar fields in a background dominated by the BH is not in contradiction with the famous no-hair theorem involving classical black hole. The no-hair theorems applied to 4d black holes can be followed from
the fact that all degrees of freedom related to the SBH are computed
from surface integrals of massless sates in 4d. Values of non-zero
scalar fields are fixed at the horizon through the so-called
Attractor Mechanism \cite{Bellucci:2007ds}. Finally notice that since there are not extra fluxes, specially NS-NS fluxes, all $D3$-branes are free from FW anomaly. This becomes an important restriction in the construction of black holes in a background threaded with NS-NS fluxes.\\

\section{Black holes from half-flat manifolds}
So far we have reviewed the standard construction of supersymmetric black holes  by wrapping $D3$-branes on homological cycles of a Calabi-Yau manifold.  In this section we shall concentrate our analysis in constructing black holes by wrapping $D3$-branes on a half-flat manifold. As we shall see, this implies wrapping $D3$-branes on torsional cycles and leading   to the existence of extra degrees of freedom associated to the torsional group \cite{Camara:2011jg}. \\

\subsection{Why selecting half-flat manifolds?}
As studied in the last few years, generalized CY manifolds can be characterized by torsional components of the Levi-Civita connection as representations of the internal structure group $SU(3)$ \cite{Grana:2005jc}. Under this perspective, the K\"ahler and the holomorphic $(3,0)$ form satisfies:
\begin{eqnarray}
dJ&=&\frac{3}{2}Im~(\bar{W_1}\Omega_3)+W_4\wedge J + W_3,\nonumber\\
d\Omega&=&W_1J^2+W_2\wedge J+\bar{W}_5\wedge \Omega_3,
\end{eqnarray}
where $W_i$'s are the representations on $SU(3)$ of the intrinsic torsional components of the connection $\nabla$. The intrinsic torsion ${\cal T}$ is defined as the anti-symmetrization of the contorsion $\kappa$ which in turn is given as follows. Consider differentiation of a generic $p$-cochain  $d\sigma_p=(\nabla\sigma)_{\mu_1\cdots\mu_p}dx^{\mu_1}\cdots dx^{\mu_p}$. If $\sigma_p$ is not closed under $d$, then $d\sigma_p=\kappa\sigma_p$ where $\kappa$ defines the contorsion. Then we can define a differential operator $d^{(T)}$ with torsion such that $d^{(T)}\sigma_p=0$ with
\begin{equation}
d^{(T)}=d - \kappa.
\end{equation}
In order to wrap D3-branes on cycles of the internal manifold $Y_3$ we concentrate on those manifolds where the non-vanishing terms of $dJ$ and $d\Omega$ have the following two properties: 1) Torsional components of the connection are represented by torsional components of the (co)homology such that it is still possible to wrap $D3$-branes in a geometrical way, and 2) we need to relate the $SU(3)$ representations of the cohomology groups of $Y_3$ with the $SU(3)$-representations of the intrinsic torsion. This forces us to consider the case in which $W_4=W_5=0$ since the corresponding cohomology groups vanish in a CY manifold.\\

There is a variety of manifolds for which  $W_4=W_5=0$ such as Calabi-Yau, Almost K\"ahler, Nearly K\"ahler, Special hermitian and Half-Flat \cite{Grana:2005jc}. We are going to focus on the simplest and more studied case of the half-flat manifold, which torsional cohomology components are easy to compute and therefore a detailed study of how $D3$-branes wrap such components can be carried out straightforwardly. Nevertheless, it is important to mention that by selecting a half-flat manifold as a background to built black holes, some extra effects (with respect to the standard supersymmetric black hole in a Calabi-Yau) would come precisely from torsional branes, by which we mean $D3$-branes wrapping torsional cycles. \\

\subsection{Half-flat manifolds}
Let us start by reviewing the construction of the cohomology groups associated to a half-flat manifold $Y_3$.
Under mirror symmetry  compactification of Type IIA string theory
on a CY manifold $X_3$ threaded with electric NS flux is mapped into a mirror
manifold $Y_3$ referred to as a half-flat manifold \cite{Gurrieri:2002iw,
KashaniPoor:2006si} on which Type IIB is compactified. Mirror symmetry is guaranteed once we have that on $Y_3$, $dIm~\Omega_3=0$ and $d Re~\Omega_3=e_i\wt\omega_i$, where $e_i$ comes from turning on the electric part of the NS-NS field strength in Type IIA compactification, while $\widetilde{\omega}_i$ are the 4-forms in $\text{H}^4(Y_3;\mathbf{Z})$. Here we want to stress that this fact leads to the existence of torsional components in the (co)homology of $Y_3$ as shown in   \cite{Tomasiello:2005bp, Marchesano:2006ns}.\\

Take the zero-components of a symplectic 3-form basis
$(\alpha_I, \beta^I)$ with $I=0,\dots, h^{(2,1)}(Y_3)$ satisfying
\begin{eqnarray}
d\alpha_0&=&e_i\tilde{\omega}^i\label{alpha},\\
d\omega_i&=&e_i\beta^0\label{beta},
\end{eqnarray}
where $i,j=1,\dots , h^{(2,1)}(Y_3)$. \\

Writing the right-hand side of Eq.(\ref{alpha}) as
$e_i\tilde\omega^i= k( n_i\tilde\omega^i)$ where $k=
gcd(e_1,\cdots,e_{h^{1,1}})$ for some integers $n_i$, a basis 
for $\H^{(2,2)}(Y_3;\mathbf{Z})$ is then given by
\begin{eqnarray}
\left( n_1\tilde\omega^1+ n_a\tilde\omega^a, \tilde\omega^a\right)
\end{eqnarray}
with  $a=2,...,h^{(1,1)}$. It is clear that $n_1\tilde\omega^1+ n_a\tilde\omega^a$ is torsional since
$k(n_1\tilde\omega^1+ n_a\tilde\omega^a)=d\alpha_0$,
but $\tilde\omega^a$ is not. Hence,
\begin{eqnarray}
\H^{(2,2)}(Y_3;\mathbf{Z})=\mathbf{Z}^{h^{(1,1)}-1}\oplus
\mathbf{Z}_k
\end{eqnarray}

Following the notation used in \cite{Camara:2011jg}, we shall denote by $\wh\Omega^p(Y_3)$ all those non-closed $p$-forms such that $d\sigma_p=k\lambda_{p+1}$, implying in turn that $\lambda_{p+1}\in Tor~\H^p(Y_3;\mb{Z})$.
Therefore, from Eq. (\ref{beta}) we observe that  2-forms $\omega_i$ are non-closed under differentiation and that $d(n^i\omega_i)=  k(n^in_i\beta^0)$,
implying that $[n^in_i\beta^0]\equiv \beta^{0,tor}\in {\rm Tor}~\H^3(Y_3)$ and $n_i\wt\omega_i \in \wh\Omega^2(Y_3)$  (we have taken $n^in_i=1$) . Hence
there is  a single 3-form which is torsional ($\beta^{0,tor}$) and another which
is non-closed ($\alpha_0\equiv\wh\alpha_0$). From this it is concluded that
\begin{eqnarray}
\H^3(Y_3;\mb{Z})=\mathbf{Z}^{2h^{(1,2)}}\oplus \mathbf{Z}_k.
\end{eqnarray}
With respect to the 2-forms we can construct a basis of $\H^{2}(Y_3;\bf{Z})$ given by
\begin{eqnarray}
\left(\omega_1,  \eta_a=\omega_a - \frac{e_a}{e_1}\omega_1\right).
\end{eqnarray}
The forms $\eta^a$ are all closed, but $\omega_1$ is not. Notice
also that none of them (including $\omega^1$) are torsional. Hence
\begin{eqnarray}
\H^{(1,1)}(Y_3;\mb{Z})= \mathbf{Z}^{h^{(1,1)}-1}.
\end{eqnarray}
The results are summarized in Table 1.\\

\begin{table}
\begin{center}
\begin{tabular}{|c|cccc|}
\hline
& ${\text{H}}^n({Y}_3;\mathbf{Z})$&${\rm Tor}~{\rm H}^n(Y_3)$&exact $mod$ $k$&non-closed\\
\hline
$n=0$&$\mathbf{Z}$&--&--&--\\
$n=1$&--&--&--&--\\
$n=2$&$\mathbf{Z}^{h^{(1,1)}-1}$&--&--&$n^i\omega_i\equiv\wh\omega_2$\\
$n=3$&$\mathbf{Z}^{2h^{(2,1)}}$&$\mathbf{Z}_k$&$n^in_i\beta^0\equiv\beta^{0, tor}$&$\wh\alpha_0$\\
$n=4$&$\mathbf{Z}^{h^{(1,1)}-1}$&$\mathbf{Z}_k$&$n_i\tilde{\omega}^i\equiv\omega_4^{tor}$&--\\
$n=5$&--&--&--&--\\
$n=6$&$\mathbf{Z}$&--&--&--\\
 \hline
\end{tabular}
\caption{Cohomology groups for $Y_3$.}
\label{tab:a}
\end{center}
\end{table}

The existence of torsional forms in a given manifold and their closure under the action of the Laplacian \cite{Marchesano:2006ns} leads to the fact that it is possible to expand 
 RR potentials in terms of  forms belonging to $Tor~\H^p(Y_3,\mathbf{Z})\oplus \widehat{\Omega}(Y_3)$. Notice that under differentiation
\begin{eqnarray}
d&:&\widehat{\Omega}^p(Y_3)\rightarrow \text{Ker}\left[Tor~\H^{p+1}(Y_3;\mathbf{Z})\right].\\
\end{eqnarray}
Therefore, following the notation used in \cite{Camara:2011jg} it is possible to define a basis of 3-forms in the half-flat manifold as $(\wh{\alpha}_0,\beta^{0,tor})$ supported in the pair  $(\Sigma_3^{tor}, \wh{\Pi_3})$ conformed by a 3-cycle and a 3-chain with
\begin{eqnarray}
k\Sigma_3^{tor}&=&\partial \wh{\Pi}_4,\nonumber\\
\partial\wh\Pi_3&=&k\Sigma_2^{tor},
\end{eqnarray}
where $\Sigma_3^{tor}\in Tor~\H_3(Y_3;\mb{Z})$ and $\wh\Pi_3\in\wh\Omega_3(Y_3)$. This establishes  an isomorphism between the spaces $Tor~\H^3(Y_3)\oplus\wh\Omega^3$ and $\wh\Omega_3\oplus Tor~\H_3(Y_3)$, meaning that the trivial element in the field is given by integration of a torsional (non-closed) form over a non-closed (torsional) cycle. This in turn defines an extra isomorphism between $Tor~\H^3(Y_3)$ and $Tor~\H^4(Y_3)$ as expected by Poincar\'e duality and the Universal Coefficient Theorem \cite{Bott-Tu, Camara:2011jg}. Specifically we have that
\begin{eqnarray}
\int_{\Sigma_3^{tor}}\wh{\alpha}_0&=&-\int_{\wh\Pi_3}\beta^{0,tor}=\int_{Y_3}\wh\alpha_0\wedge\beta^{0,tor}=1,\nonumber\\
\int_{\wh\Pi_4}\omega_4^{tor}&=&-\int_{\Sigma_2^{tor}}\wh\omega_2=\int_{Y_3}\omega_4^{tor}\wedge\wh\omega_2=1,
\end{eqnarray}
in accordance with the basis chosen in \cite{Louis:2002ny, Gurrieri:2002wz, Gurrieri:2002iw, KashaniPoor:2006si} and where we  made use of
\beq
\int_{\Sigma_3^{tor}}\wh\alpha_0=\frac{1}{k}\int_{\partial \wh\Pi_4}\wh\alpha_0=\frac{1}{k}\int_{\wh\Pi_4} d\wh\alpha_0=\int_{\wh\Pi_4}\omega_4^{tor}=1.
\eeq
\\
From these relations we can also obtain that
\begin{eqnarray}
{\cal PD}_6(\wh\alpha_0)&=&\wh\Pi_3,\nonumber\\
{\cal PD}_6(\beta^{0,tor})&=&\Sigma_3^{tor},
\end{eqnarray}
with ${\cal PD}_6: Tor~\H^3(Y_3;\mathbf{Z})\oplus\wh\Omega^3(Y_3) \longleftrightarrow Tor~\H_3(Y_3;\mathbf{Z})\oplus \wh\Omega_3(Y_3)$. Notice that the above integrals define the linking number between $\S$ and $\P$.\\

Using this structure it is possible to write down expressions for the K\"ahler form and the holomorphic 3-form depending on the non-cohomological forms \cite{Gurrieri:2002iw, KashaniPoor:2006si} .
Consider the K\"ahler 2-form $J = v^i \omega_i$, which can be written in terms of $\wh\omega_2$ as
\begin{equation}
J=v^in_i\wh\omega_2 ,
\label{J}
\end{equation}
from which it follows that
\begin{equation}
d J = v^i n_id \wh\omega_2 = v^i n_i k\beta^{tor, 0}, \qquad n_i \in \mathbb{Z}.
\end{equation}
Similarly, the holomorphic $(3,0)$-form $\Omega_3$ satisfies \cite{Gurrieri:2002iw, KashaniPoor:2006si} 
\begin{equation}
d\Omega=d\wh\alpha_0=k\omega_4^{tor},
\end{equation}
for which it is straightforward to set the most general expression for $\Omega_3$:
\begin{equation}
\Omega_3 =\Omega^0_3+ \wt\Omega_3=X^i\alpha_i-F_i\beta^i+\wh\alpha_0 -F_0\beta^{0,tor}, 
\label{omega}
\end{equation}
with $\wt\Omega_3$ corresponding to the components of $\Omega_3$ expanded in the basis $(\wh\alpha_0, \beta^{0,tor})$ and where the periods are given  by the integrals
\begin{eqnarray}
F_I=(F_0, F_i)&=&\left(\quad\int_{\wh\Pi_3}\wt\Omega_3\;,\int_{B_i}\Omega^0_3\quad\right),\nonumber\\
X^I=(X^0, X^i)&=&\left(\quad\int_{\Sigma_3^{tor}}\wt\Omega_3\;,\int_{A^i}\Omega^0_3\quad\right).
\end{eqnarray}

Notice that for the half-flat manifold, the non-closed parts of  $J$ and $\Omega$ parametrize how different a half-flat manifold is compared with a CY manifold.  Particularly, for the half-flat,  the torsional components of the geometrical connection are identified with torsional components of cohomology. This is a key ingredient in our method to construct BH by wrapping $D$-branes on internal cycles, since the extra information we have in relation with a CY manifold is now encoded in torsional homology cycles, which one can use to wrap $D$-branes. Notice that this is just the half-flat version of the well known example in which NS-NS flux is transformed into torsional cohomology at the level of the tori compactification  \cite{Marchesano:2006ns}.\\

\subsection{Discrete electric gauge charge from half-flat manifolds}
In contrast with the supersymmetric compactification on a CY manifold,  a half-flat manifold has torsional cycles. Following the prescription reviewed in Section 2, one wonders what are the consequences of wrapping $D3$-branes around some of these spaces on the black hole physics. It is then the purpose of this section to study the physical implications of wrapping branes on torsional 3-cycles. For that, as we have seen we must in principle also consider chains in $\wh\Omega_3(Y_3)$.\\

Let us start by wrapping $N$ $D3$-branes on  a general chain $\wt{\cal C}_3 \in  Tor~\text{H}_3(Y_3;\mathbf{Z}) \oplus \wh\Omega_3(Y_3)$ given by
\beq
\C=p^0\P-q_0\S,
\eeq
with a worldvolume of the $D3$-branes  given by ${\cal W}_4=\gamma\times\C$. It follows that the electric charge is computed by
\beq
Q_3=\int_{\Gamma_6}{\cal PD}({\cal W}_4)=\int_{\Gamma_6}{\cal PD}_4(\gamma)\wedge{\cal PD}_6(\C).
\eeq
Contrary to the SBH in which $\Gamma_6=\mathbf{B}^3\times \Gamma_3$ with $\Gamma_3\in \H_3(Y_3;\mathbf{Z})$ in this case we can capture a discrete charge value by integrating the current ${\cal PD}_6(\C)$ over the chain
\beq
\Gamma_6\equiv\mathbf{B}^3\times\wt\Gamma_3=\mathbf{B}^3\times\left(\frac{e^0}{k}\P-m_0\S\right), 
\eeq
which is nothing else that the worldvolume of a $D3$-brane wrapping the torsional 2-cycle
\beq
\partial^2\Gamma_6=\mathbf{S}^2\times e^0\Sigma_2^{tor},
\eeq
precisely corresponding to the fractional charge computed by the Aharanov-Bohm effect through the holonomy of a 4d-string around the point-like BH constructed by wrapping $D3$-branes on $\C$ \cite{Wen:1985qj, Preskill:1990ty, Coleman:1991ku,  Banks:2010zn, Camara:2011jg, BerasaluceGonzalez:2012zn}. Therefore, it follows that
\begin{eqnarray}
*F_3&=&{\cal PD}_6(\C)=p^0\wh\alpha_0-q_0\b, \nonumber\\
F_3&=&{\cal PD}_6(\G)=\frac{e^0}{k}\a-m_0\b,
\label{F3}
\end{eqnarray}
with
\begin{eqnarray}
p^0&=&\frac{e^0}{k}A^0_0-m_0C^{00},\nonumber\\
-q_0&=&m_0A^0_0+\frac{e^0}{k}B_{00},
\end{eqnarray}
and the real  matrix elements given by
\begin{eqnarray}
A_0^0&=&-\int \a\wedge*\b,\nonumber\\
B_{00}&=&\int \a\wedge *\a,\nonumber\\
C^{00}&=&-\int \b\wedge*\b.
\label{matrices}
\end{eqnarray}
Thus the RR field strength associated to those $D3$-branes is then given by
\beq
F_5=F_2\wedge \left[\frac{e^0}{k}\a-m_0\b\right],
\label{F5D3}
\eeq
from which it is straightforward to compute the effective charges. However, 
before computing the corresponding electric and magnetic charges it is worth mentioning that,  as shown in \cite{Gurrieri:2002wz, Gurrieri:2002iw} for the half-flat compactification,  is also necessary   to consider the presence of a non-trivial NS-NS flux\footnote{Notice that although we have a NS-NS flux, the low energy limit preserves a ${\cal N}=2$ supersymmetry since we are not considering an extra RR flux $F_3$ and therefore the tadpole contribution to $D3$-brane charge vanishes \cite{Taylor:1999ii}.} given by $H_3= e_0\beta^{0,tor}$. Nevertheless,  the existence of this flux is potentially dangerous for  $D3$-branes wrapping regions on which the NS-NS flux is supported, since it renders the branes anomalous 
\cite{Freed:1999vc, Maldacena:2001xj, Evslin:2006cj} . In order to cancel this Freed-Witten anomaly it is necessary that
\beq
\int_{\wt{\cal C}_3}e_0\beta^{0,tor}=0.
\eeq
implying that $p^0=0$. Therefore,  Freed-Witten anomaly cancelation leads us to a relation between the winding numbers $e^0$ and $m_0$ by
\beq
m_0=\frac{e^0}{k}\frac{A}{C},
\label{FW}
\eeq
where we have adopted the notation of $A, B$ and $C$ to refer to the corresponding matrix elements in Eq. (\ref{matrices}). Let us emphasize 
two important remarks: 
\begin{enumerate}
\item
For $e^0=k$ the worldvolume of the 4-dimensional string becomes trivial, and no measurement of fractional charge is obtained. Therefore the value of the quotient $e^0/k$ vanishes if it equals an integer, i.e. we must refer to it as $e^0/k~mod ~1$.
\item
By canceling  the Freed-Witten anomaly,  the internal 3-form $F_3$ reduces its degrees of freedom from 2 to 1.  By writing $F_5$ as 
\beq
F_5= e^0F_2\wedge \wh\alpha_0 - m_0F_2\wedge \beta^{0,tor}= F^0\wedge\wh\alpha_0- G_0\wedge\beta^{0,tor},
\eeq
it is possible to eliminate $G_0$, since it does not carry degrees of freedom. This actually was shown in \cite{Gurrieri:2002wz, Gurrieri:2002iw} by compactifying Type IIB string theory on a half-flat manifold and by demanding self-duality on the 5-form field strength. Therefore, it seems that self-duality on  $F_5$ is in agreement with the cancelation of Freed-Witten anomaly on $D3$-branes wrapping $\P$. Notice  that this implies that the chain $\C$ reduces to a torsional cycle, i.e., $D3$-branes are only wrapping torsional components in the homology of $Y_3$. After making use of the FW anomaly cancelation, $\C$ and $\G$ reduce to
\begin{eqnarray}
\C&=&-\left(\frac{e^0}{k}~mod~1\right)\frac{1}{C}\S,\nonumber\\
\G&=&\e\left[\P-\frac{A}{C}\S\right],
\end{eqnarray}
and
\beq
F_5=\e F_2\wedge (\a-\frac{A}{C}\b).
\eeq
\end{enumerate}

We have now all the necessary ingredients to compute the black hole  electric and magnetic charges,  which read
\begin{eqnarray}
Q_e&=&Q\int_{\G}\ast F_3= \frac{Q}{C}\e^2,\nonumber\\
Q_m&=&P\int_{\C}F_3=-\frac{P}{C}\e^2,
\end{eqnarray}
where we have used that the effective charges are 
\beq
Q=\int_{\mathbf{S}^2}\ast F_2, \qquad \text{and} \qquad P=\int_{\mathbf{S}^2} F_2.
\eeq

The total charge can also be computed by integrating the self-dual 5-form $\mathscr{F}_5=F_5+\ast_{10} F_5$ over the cycle ${\cal W}_5$ given by ${\cal W}_5= \mathbf{S}^2\times (\G\cup\C)$, and it is given by
\begin{eqnarray}
Q_{TOT}=\int_{\Gamma_5}\mathscr{F}_5= \frac{1}{C}(Q-P)\e^2,
\end{eqnarray}
where we have used the Freed-Witten anomaly cancellation condition (\ref{FW}).  Notice that once we have wrapped $k$-$D3$-branes on $\C$, their worldvolume becomes trivial in homology and in consequence,  $Q_{TOT}$ vanishes. This is exactly the mirror symmetric picture of the disappearance of $D$-branes in a background threaded with NS-NS flux with support on the homology cycles on which the $D$-branes are wrapped \cite{Danielsson:2006jg, LoaizaBrito:2007kz}.\\

Under this perspective, measuring a discrete charge by an Aharanov-Bohm mechanism through the presence of a 4d-string, indicates the existence of extra degrees of freedom associated to the black hole as pointed out in \cite{Coleman:1991ku, Camara:2011jg}. Therefore, by considering a small number $k$ and a huge number of $D3$-branes wrapping torsional cycles in $Y_3$ the effects of torsional $D3$-branes are manifested at the quantum level. Besides this, it is also possible to show that at the low energy level, there are massive scalars which are charged under the graviphoton, with a discrete charge as well (see Appendix B).  In conclusion, all together, these features point out to the presence of the so called {\it quantum hair} of a black hole, which in our case is supersymmetric.\\

Before computing some extra consequences in the mass of the black hole, let us remark one last comment concerning the electric charge we have computed from torsional branes:

\begin{enumerate}
\item
We know  that under mirror symmetry, the  electric component of a NS-NS flux is mapped into the geometry of a manifold called {\it half-flat}. Now, by considering the construction of SBH in such backgrounds we can safely say that a SBH constructed in a CY manifold threaded with electric NS-NS flux is mapped into the mirror symmetric picture in which a SBH has {\it quantum hair}. 
\item
These discrete charges are associated to massive gauge bosons, which
obtain their masses by the breakdown of a continuous symmetry $U(1)$
into a discrete  $\mathbf{Z}_k$ (for details see
\cite{Coleman:1991ku, Camara:2011jg}). In consequence, these massive
gauge fields are relevant at the scale just below  the
breaking of symmetry\footnote{ For a string scenario in which discrete symmetries arise by the rupture of a continuous symmetry see \cite{Berasaluce-Gonzalez:2013sna}.}
\end{enumerate}

\subsection{Corrections of Black Hole Mass}
Up to now we have computed the  discrete charge of a SBH related to $D3$-branes  wrapping a torsional 3-cycle  $\C$. It is therefore of importance to compute the contribution to the mass  of the BH given by  those  $D$-branes.  As it has been remarked, after using $k$ $D3$-branes, the 3-cycle $\C$ becomes trivial in homology and collapses into a point, rendering all the involved $D3$-branes to become unstable, canceling out all their RR charge and transforming into closed strings \cite{Maldacena:2001xj, Evslin:2006cj, Marchesano:2006ns}. It is therefore expected to associate a discrete value of the mass. An important point to remark is that while accumulating $D3$-branes the system is stable and behaves as a BPS object\footnote{Notice that this is  valid since in this case the charge  computed through $dJ$ over a chain in $\wh\Omega_3(Y_3)$  does not vanish.} in the 4-dimensional extended space.\\

Hence, with the purpose of computing the mass from a base point of view, let us closely follow the black hole's mass computation given in \cite{Denef:1999th}. Let us start by considering the Dirac-Born-Infeld (DBI) action of a bunch of $D3$-branes wrapping $\C$ given by\footnote{We are taking all  numerical coefficients equal to 1.}
\begin{equation}
S_{DBI} = -\int_{{\cal W}_4} \sqrt{-h}\ast\mathbf{1},
\label{one}
\end{equation}
where $h$ is the determinant of the pull-back of the 10-dimensional metric to the worldvolume ${\cal W}_4=\gamma\times\C$ of the $D3$-branes considered  stable unless we have a number of $k$ $D3$-branes. Therefore, let us take only a number $e^0<k$ of $D3$-branes wrapping $\C$.\\

Under this assumption, all our branes are  stable and therefore are wrapping a chain which minimize their energy. In such geometric regions, it is possible to show \cite{Becker:1995kb, KashaniPoor:2006si} that two conditions are hold: i) $J^\ast=0$ on $\C$ and ii) the superpotential has a constant phase. These two properties lead to the possibility to write the DBI action as
\begin{eqnarray}
S_{DBI} &=&-\int_\gamma {\cal V}_{D3}\ast \mathbf{1},
\label{thirteen}
\end{eqnarray}
where ${\cal V}_{D3}$ is the volume of $D3$-branes playing the role of the 4-dimensional mass $M_{BH}$, which  in terms of the holomorphic 3-cochain $\Omega_3$, reads
\begin{eqnarray}
\label{mass}
M_{BH} &=&  e^{\K/2} \left| ~\int_{{\cal C}_3\cup\C} \Omega_3~ \right|\nonumber\\
&=& e^{\K/2}\left|~\int_{{\cal C}_3}\Omega_3^0+\int_{\C}\wt\Omega_3~\right|.
\end{eqnarray}
Therefore, using the expression (\ref{omega}) for  the torsional component of $\Omega_3$, the mass term given by the $D3$-branes wrapping $\C$ which actually is the mass contribution to the SBH by adding torsional $D3$-branes, reads
\begin{eqnarray}
\Delta M=e^{\K/2}~\int_{\C}\wt\Omega=-e^{\K/2}\e\frac{1}{C},
\end{eqnarray}
Hence, the total mass of the black hole conformed by $D3$-branes wrapping a 3-cycle ${\cal C}_3$ in $\text{H}_3(Y_3;\mathbf{Z})$ and by $D3$-branes wrapping the torsional cycle  $\C$ is  given by
\beq
M^2_{BH}=(M_{BPS}+\Delta M)^2= e^{\K}\left| m_{BPS}-\e\frac{1}{C}\right|^2,
\eeq
with $m_{BPS}$ given by (\ref{W}) as
\beq
m_{BPS}=-\mathscr{W}=-(e_iX^i-m^iF_i).
\eeq
Notice that here, for the half-flat case, the index $i$ in the superpotential runs from 1 to $h^{(1,1)}(Y_3)$, contrary to the superpotential in a CY manifold where the index also takes the zero value. Therefore, in analogy we can write the black hole total mass in terms of a new superpotential given by
\begin{eqnarray}
\mathscr{W}_{TOTAL}&=&\mathscr{W}+\int_{Y_3}\ast F_3\wedge\wt\Omega_3,\nonumber\\
&=&\mathscr{W}+\lambda\mathscr{W}_{HF},
\end{eqnarray}
where $\lambda=i\e\frac{1}{C}\frac{1}{kn_iv^i}$, and
\beq
\mathscr{W}_{HF}=\int_{Y_3}idJ\wedge \wt\Omega_3,
\eeq
which under the flux conditions in our setup (no RR fluxes and cancellation of Freed-Witten anomaly derived from the presence of non-trivial a NS-NS flux) is precisely the superpotential related to a half-flat manifold as shown in \cite{Gurrieri:2002wz, Gurrieri:2002iw}. Two comments are given in order:  First, notice that by demanding that the black hole mass satisfies the relation $M_{BPS}^2=e^\K|\W|^2$ where $\W$ is a superpotential,  provides an alternative way to derive the superpotential of the half-flat manifold. Second,  we see that the contribution to the mass by torsional branes is also proportional to $\e$, indicating that after wrapping $k$ $D3$ in $\C$, the extra mass term vanishes.

\subsection{ Lost of Quantum Hair}

As shown below, quantum degrees of freedom, or quantum hair can be  associated to the black hole by wrapping $D3$-branes on torsional cycles.  Besides this electric discrete charge we have also computed the mass contribution and see that it also has a discrete value, meaning that upon completion of $k$ $D3$-branes wrapping $\C$, the mass of the black hole will collapse to the originally value  $M_{BPS}$ (i.e., without considering torsional  branes) and it will  loose all its quantum hair.\\

This is quite interesting since it implies that a stable and extremal black hole with an associated vanishing temperature would emit some radiation (consisting on closed strings) once the number of torsional branes reach $k$. Once the SBH loose all its quantum hair it would return to another stable state with a lower mass. Therefore there must be an emission of closed strings  localized in time and in consequence we expect a variation in the entropy by the lost of all torsional degrees of freedom.  A previous mirror symmetric picture of this mechanism was partially studied in \cite{LoaizaBrito:2007kz}.\\

Hence, let us compute the change in entropy and the associated temperature for the radiation the SBH would emit once completing $k$-torsional $D3$-branes and let us start by reviewing the way in which entropy is computed in a CY manifold. The associated entropy for a SBH constructed on a CY manifold is computed by extremizing the action \cite{Ferrara:1996um, Ooguri:2005vr}
\begin{equation}
S=-\frac{\pi}{4}\left[ e^{-\mathscr{K}(X,\ov{X})} +2i\mathscr{W}(X) -2i\ov{\mathscr{W}}(\ov{X})\right],
\label{Eq:action}
\end{equation}
and evaluating the extreme at the attractor point on which the involved superpotential vanishes rendering the system supersymmetric. The K\"ahler potential is
\beq
\K=\K_0+\wt{\K}=i~log\left(\int~\Omega^0\wedge\bar{\Omega}^0+\int \wt{\Omega}_3\wedge\bar{\wt{\Omega}_3}\right)
\eeq
with $\wt\K=-\frac{1}{2}Im~X_0F^0.$ Notice that although at this point we are not considering the presence of torsional $D3$-branes, the K\"ahler potential contains some information coming from torsional cohomology since it is related to the geometry of the internal space independently of the presence of $D3$-branes. Since the supersymmetric black hole constructed by wrapping  $D3$-branes on 3-cycles satisfies the BPS bound, i.e., it is an extremal black hole, it does not radiates since it is in a state of minimal energy. Therefore the associated temperature is zero (see \cite{Dall'Agata:2011nh} and references therein).\\

Now, let us think on a system consisting on just $D3$-branes wrapping supersymmetric 3-cycles on which we start adding torsional $D3$-branes, taking care that the number of these branes does not overpass $k$.  Since we are adding extra degrees of freedom (parametrized by $e^0$) it is expected that entropy grows with respect to the entropy associated to the SBH. Its variation must come precisely from the extra components in the superpotential denoted by $\lambda\mathscr{W}_{HF}$, i.e., the entropy variation can be computed by extremizing the action
\beq
\wt{S}=\frac{\pi}{4}\left(e^{\wt\K}-4Im~\lambda\W_{HF}\right).
\eeq
A direct calculation shows that $\wt{S}$ has an extreme at 
\beq
X^0_{min}=\frac{i}{C^{00}}\e\frac{1}{(Im~\tau)_{00}}
\eeq
at which, upon substitution, gives the entropy associated to torsional branes (and by taking $\mathscr{W}_{HF}(X^0_{min})=0$):
\beq
\Delta{\cal S}=\frac{\pi}{2}\frac{1}{C^{00}}\frac{1}{Im~\tau_{00}}\e^2=\frac{\pi}{2}e^{-\K/2}\e\Delta M.
\eeq
Therefore, the entropy related to  quantum hair goes like $\e^2$ meaning that by increasing the number of torsional branes, the entropy of the system also becomes larger. Once we add $k$ $D3$-branes, the system becomes unstable to decay into the original supersymmetric setup and all torsional branes radiate into closed strings. Although a precise description of this transition is beyond the scope of this work, we can mention some interesting features. \\

First of all, by reaching the number $k$ of $D3$-branes, the extra mass and entropy vanishes.
The transition consists on a black hole which suddenly looses part of its mass and goes from a stable state with a zero-temperature to another stable state with  a smaller mass. During this transition, the system is not represented by a BPS state since $\C$ is a trivial cycle and collapses into a point. It is then natural to associate a temperature related to the emission of the energy contained in the system of $D3$-branes wrapping a trivial cycle, which being a non-supersymmetric and unstable state, can be estimated from $dS/dM=1/\Delta T$. Therefore
\beq
\Delta T\sim \frac{\pi}{2}e^{-\K_0/2}.
\eeq
From this we can also notice the following: consider two black holes with the same total mass, but one has a larger amount of mass coming from supersymmetric $D3$-branes. Therefore, the mass contribution from torsional branes is smaller in the first black hole than in the second one. In that sense, the black hole with more discrete charge is also the one which more entropy. Notice that if these black hole would be non-extremal, we would said that a black hole with more discrete charge would be also cooler. These features
are pretty similar to the properties one expects (in a supersymmetric point  of view) from a black hole with quantum hair with an associated temperature, as predicted in \cite{Preskill:1990ty} .\\

Finally, from this supersymmetric construction still there is a question we can address and that was already pointed out in \cite{Coleman:1991ku}. It would be desirable to compute the discrete charge of a black hole without considering the presence of a 4-dimensional string. We consider that this can be accomplished by the use of K-theory.

\section{Quantum hair and K-theory}

The discrete electric charge computed in the previous sections relies on the presence of an extra object. Therefore, quantum hair seems to be detectable only if we could take into account a 4-dimensional string and perform an holonomy around the black hole. This of course triggers a question about how to compute such discrete charge without using an extra extended object.   In the context of string theory, computation of classical properties of a SBH requires the use of (co)homology, while the  quantum regime of the black hole should be described in a appropriate way related to the computation of $D$-brane RR charges. From some years, we know that such mathematical structure is encoded in K-theory and for that reason we expect that discrete charges
must be derived from some version of K-theory. In this section we shall use the Atiyah-Hirzebruch Spectral Sequence (AHSS), connecting cohomology to K-theory in order to derive how the discrete charge appears by computing the corresponding K-theoretical charge of the $D$-branes used for the construction of the black hole.\\

With the purpose of presenting a clearer argument and in order to show that SBH can also be constructed by compactifying Type IIA string theory on a half-flat manifold, we shall present our analysis in this background, i.e., the construction of black holes by wrapping $D2$- and $D4$-branes in Type IIA theory compactified on a half-flat manifold.

\subsection{The Atiyah-Hirzebruch Spectral Sequence}
Let us start by briefly reviewing the AHSS in the context of string theory.
Essentially the AHSS is an algorithm which connects integral
cohomology to  K-theory \cite{Diaconescu:2000wy,
Bergman:2001rp, GarciaCompean:2002kc}. The main goal of this
approach is to compute  the K-theory group $K(X)$ related to the RR charge of $D$-brane supported on the submanifold $X$ with dimension $d$. For that, the AHSS  makes use of a sequence of successive
approximations starting from integral cohomology and gradually considering 
successive
orders of approximation which involves the cohomology of differential maps
$d^n$, where $d^n:H^p(X;\mathbf{Z})\rightarrow H^{p+n}(X;\mathbf{Z})$.  In each
step, the $n$-cohomology group $E_p^n$ for a given  $n$ is computed by
the quotient $K_{p}(X)/K_{p+1}(X)$ where $K_p(X)$ is a subgroup of
$K(X)$ which classifies all stable $D(d-p)$-branes supported on a
$(d-p)$-dimensional submanifold of $X$ but  trivial in
$(d-p-1)$-submanifolds via the RR field strengths ($p$-forms). Computing $K_p(X)$ involves solving the following exact short sequence:
\beq
\diagram{0&\harr{}{}&K_{p+1}&\harr{}{}&K_p&\harr{}{}&K_p/K_{p+1}&\harr{}{}&0}\;.
\label{ext}
\eeq
If all extensions are trivial for all $p$, the K-theory group $K(X)$
is computed just by adding the subgroups $E_p^n$, i.e., by $K(X)=\oplus_p E^n_p$. Notice therefore, that in a fluxless CY compactification $D$-brane charges or equivalently RR charge is simply computed through the cohomology groups. By turning on an extra NS-NS flux, the AHSS requires a second step of approximation involving the groups $E_p^3$ (\cite{LoaizaBrito:2004hk, Bergman:2001rp, Evslin:2006cj}). Hence, in the absence of extra fluxes (notice that $H_3=e_0\b$ does not have an influence in the sequence), $K(X)=H^p(X)$ up to solving the extension problem (\ref{ext}). \\

\subsection{Discrete charge from K-theory}

A SBH in the context of Type IIA string theory is constructed by wrapping $D2$ and $D4$-branes on 2- and 4-dimensional  chains in $Y_3$. Therefore the electric and magnetic charges are computed by integrating the RR field strength (a,  4- and 6-form, respectively) over some suitable submanifold of the ten-dimensional space-time, this is
\beq
Q_e^{IIA}=\int_{\mathbf{S}^2\times\Gamma_4}\ast F_2\wedge {\cal PD}_6({\cal C}_2), \quad\text{and}\quad Q_m^{IIA}=\int_{\mathbf{S}^2\times\Gamma_2}F_2\wedge{\cal PD}_6({\cal C}_4),
\eeq
with both currents ${\cal PD}_6({\cal C}_2)$ and ${\cal PD}_6({\cal C}_4)$ in $\H^4(Y_3;\mathbf{Z})$ and $\H^2(Y_3;\mathbf{Z})$ respectively. 
After extending the integration to the whole internal space, 
the charge of a BH in four-dimensions is determined by a 6-form
flux in $\H^6(Y_3;\mathbf{Z})$, proportional to ${\cal PD}_6({\cal C}_2)\wedge J$ for a $D2$-brane on ${\cal C}_2$ and ${\cal PD}_6({\cal C}_4)\wedge J^2$ for a $D4$-brane wrapping ${\cal PD}_6({\cal C}_4)$. Since it is not obvious from the above expressions that the
corresponding charges are fractional without considering  the
presence of a 4d string, our goal here is to elucidate its nature from the K-theory perspective.\\

Within the context of the AHSS, the relevant short sequence involves
the following filtrations (where $h^{(1,1)}$ is the Hodge number of $Y_3$):

\begin{enumerate}
\item
 $K_5=K_6=\mathbf{Z}$ which measures the charge 
carried by the 6-form ${\cal PD}_6({\cal C}_2)\wedge J_2$ by computing the cohomology group $\H^6(Y_3;\mathbf{Z})=\mathbf{Z}$.
\item
 $K_4=\mathbf{Z}\oplus\mathbf{Z}^{h^{(1,1)}}$ which measures the K-theoretical charge related to the 6-form ${\cal PD}_6({\cal C}_2)\wedge J_2$ and
stable $D2$-branes wrapping  $h^{(1,1)}$ 2-cycles in $Y_3$, and
\item
$K_4/K_5=H^4(Y_3;\mathbf{Z})=\mathbf{Z}^{h^{(1,1)}} \oplus
\mathbf{Z}_k$ concerning the group of 4-forms,  related only to $D2$-branes wrapping 2-cycles of $Y_3$. Notice the presence of torsional components for the half-flat manifold.
\end{enumerate}

Therefore,
the relevant extension problem is given by
{\small
\begin{eqnarray}
\diagram{
 0&\harr{\times k}{}&H^6(Y_3;\mathbf{Z})=\mathbf{Z}&\harr{}{}&\mathbf{Z}\oplus\mathbf{Z}^{h^{(1,1)}}&\harr{}{}&H^4(Y_3;\mathbf{Z})=\mathbf{Z}^{h^{(1,1)}}\oplus \mathbf{Z}_k&\harr{}{}&0\cr}\;,
\end{eqnarray}}
where we have also assumed that there is no difference between
cohomology and K-theory in the sequential steps. By demanding the
sequence to be exact we notice that there must be a shift of
fractional charge $i/k$ for each element in $\mathbf{Z}_k$, with
$i=1,\dots , k$.\\

Therefore, for each $D2$-brane wrapping a torsional cycle in
$\mathbf{Z}_k$, the corresponding torsional element induces a
fractional charge in the generator $G_6 \in \H^6(Y_3;\mathbf{Z})$ which as said, contributes to the 4-dimensional charge of the BH. Notice that
a similar situations holds
for the magnetic part, i.e., by wrapping $D4$-branes on 4-cycles and that the fractional charge induction is independently of the presence of one or another. This in fact confirms that it is possible to associate a fractional K-theoretical charge of branes wrapping torsional cycles.\\

\section{Final Comments}
In this work we have constructed a supersymmetric black hole in the effective low-energy theory by wrapping  $D3$-branes on 3-cycles of a half-flat manifold. As it is well known, Type II string compactification on half-flat manifold is the mirror symmetric image of a compactification on a Calabi-Yau manifold threaded with electric NS-flux. Therefore, we are wrapping $D3$-branes on a manifold which has back-reacted under the presence of the electric NS-flux and in consequence, it is expected that the black hole constructed in such scenario contains some characteristics inherited from the electric NS-flux.\\

Those effects manifest in the black hole's  physics, primarily by the presence of torsional components in the (co)homology of the half-flat manifold with an associated discrete group denoted by $\mathbf{Z}_k$.  A number of $N$ $D3$-branes wrapping torsional cycles correspond in the low-energy level to stable and supersymmetric point-like objects with a discrete value $N/k~mod~1$  for the mass and for the electric and magnetic charges. Expansion of the corresponding RR potential on these torsional components of cohomology leads to the existence of effective massive gauge bosons and massive scalars with discrete gaugings.\\

Since $k$ is finite, by wrapping a large number of $D3$  on non-torsional cycles,  a massive SBH is
constructed with $M_{BH}>>M_{pl}$. This is a very well approximation
of a classical SBH. However, if the number of D-branes wrapping the
non-torsional cycles are of order $k$, the massive
states related to the torsional part must become relevant. These
degrees of freedom must correspond to some hair on the SBH which
manifest in a quantum regime as {\it quantum hair} studied in
\cite{Coleman:1991ku, Banks:2010zn}.\\

Thinking on a black hole which increases its mass by adding torsional $D3$-branes it is possible to compute its variation on mass and entropy up to the non-statical stage in which the bunch of torsional branes complete the number $k$ and annihilate each other, departing from the stable BPS state. As for the electric and magnetic charges, the variation of mass and entropy resulting from increasing the number of torsional branes goes like $N/k~mod~1$. In consequence, once the number of torsional branes reach the number $k$, the black hole transits from a stable state with a mass and charge larger than a black hole conformed only by $D3$-branes wrapping homological cycles to another stable state corresponding to the supersymmetric black hole. Both states are stable, but during the transition, the system conformed by the torsional branes become an unstable set of branes wrapping a trivial cycle. Therefore, all the entropy gained during the addition of torsional branes is emitted in the form of closed strings and we can estimate a temperature related to this process. We observe that for two black holes with the same total  mass, the one with more discrete charge has a bigger entropy than the second one. This resembles some properties expected from quantum hair in a supersymmetric version.\\

Keeping the number of torsional branes less than $k$, we notice some other important features: since in a half-flat manifold there is a non-trivial NS-NS flux it is important to cancel Freed-Witten anomaly on all those branes wrapping submanifolds on which the flux is supported. This implies vanishing half of the degrees of freedom,  associated with the 5-form field strength $F_5$. This is compatible with the same lost of degrees of freedom by restricting the 5-form field strength to be self-dual as shown in \cite{Gurrieri:2002wz}. Therefore, we conclude that in half-flat compactification where $F_5$ is taken to be self-dual, all $D3$-branes are free from Freed-Witten anomalies.\\

Nevertheless, in this set up, computation of discrete quantum hair  requires the presence of a 4-dimensional string. With the purpose to compute the discrete charge of the black hole without requiring the presence of an extra object, we use the Atiyah-Hirzebruch Spectral Sequence to compute the $D3$-branes charge from a K-theoretical perspective. We find that, as in the presence of orientifolds, torsional components of cohomology induces a lift on the generators of the $D$-brane charge in fraction, rendering the total charge to be discrete.\\

However, there are still many features to study among which we can mention some: First, although it seems possible that magnetic quantum hair appears by constructing black holes on mirror symmetric manifolds to those on which magnetic components of a NS-flux have been taken into account, still it is not clear how to wrap $D$-branes on those backgrounds. Constructing black holes in general manifolds would be an important task to perform as well as the relation between those black holes and the solutions in the gauged supergravity side. Second, it would be interesting to elucidate some string mechanism which leads us to the construction of non-abelian quantum hair. Black holes in a background mirror symmetric to a compactification on a CY threaded with magnetic NS-fluxes could be the string construction of the magnetic quantum hair as described in \cite{Coleman:1991ku}. It would be interesting to explore such issue. Finally, it is necessary to study how stable is a black hole in a half-flat manifold due to the presence of an effective scalar field $V_g$ originated by the own non-zero curvature of the half-flat manifold. We leave this feature for a future work.


\begin{center}
{\bf Acknowledgements}
\end{center}
We thank Liliana Vazquez-Mercado for collaboration at the beginning of this project and to Fernando Marchesano, Andrei Micu, Gustavo Niz, Octavio Obregon, Kin-Ya Oda and Miguel Sabido for many useful discussions and suggestions. H. G.-C. is supported by the CONACyT grant 128761. O.L.-B. is partially supported by the CONACyT grant 132166 and by PROMEP under the program "red de cuerpos acad\'emicos de gravitaci\'on y f\'isica matem\'atica 2013-2014". A. M.-M. and R.S.-S. are supported by a postdoctoral PROMEP grant. 

\appendix

\section{Notation}
For a fluxless compactification on a Calabi-Yau, we have that
\begin{eqnarray}
\int_{X}\alpha_I\wedge *\alpha_J&=&\int_X \alpha_J\wedge *\alpha_I=B_{IJ},\nonumber\\
\int_X\alpha_I\wedge *\beta^J&=&\int_X\beta^J\wedge *\alpha_I=-A^I_J,\nonumber\\
\int_X \beta^I\wedge *\beta^J&=&\int_X \beta^I\wedge* \beta^J=-C^{IJ},
\end{eqnarray}
which in turn define the complex matrix ${\cal M}_{IJ}$ through
\begin{eqnarray}
A&=& (Re~{\cal M})(Im~{\cal M})^{-1},\nonumber\\
B&=& -(Im~{\cal M})-(Re~{\cal M})(Im~{\cal M})^{-1}(Re~{\cal M}),\nonumber\\
C&=&(Im~{\cal M})^{-1}.
\label{matricesap}
\end{eqnarray}

\section{Low energy theory}
In this section we review the low energy limit of Type IIB string theory compactified on a half-flat manifold closely following \cite{Louis:2002ny, Gurrieri:2002wz, Gurrieri:2002iw}.\\

First of all, it is important to notice that as shown in \cite{Camara:2011jg} all fields expanded in terms of $\wh\alpha_0$ and $\beta^{0,tor}$ are massive in the four dimensional effective theory and since $[\nabla^2, d]=0$ it is possible to show that
\beq
\nabla^2\wh\omega_2=-n_iM_i^j\omega_j=\wh\omega_2,
\eeq
where $-M^i_j=\delta^i_j$ is the corresponding mass matrix. Similarly we have that
\begin{eqnarray}
\nabla^2\wh\alpha_0&=&\wh\alpha_0,\nonumber\\
\nabla^2\beta^{0,tor}&=&\beta^{0,tor},\nonumber\\
\nabla^2\omega_4^{tor}&=&\omega_4^{tor}.
\end{eqnarray}
The squared masses are all of order of the Planck mass (we have taken $M_{pl}=1$) and therefore the Laplace operator acting on these fields gives terms of order (flux)$^2$, implying that it is possible to ignore massive KK states since the order of the fluxes is smaller that the order of the compactification scale rendering the supergravity approach valid.\\

The massive scalar fields and massive gauge vector arising from compactification on a half-flat manifold can  be shown by directly computing the effective low-energy scale as in \cite{Gurrieri:2002wz}. Let us review this computation for our specific case in which we are turning on an internal field related only with the presence of wrapped $D3$-branes, i.e., we are not considering extra NS-NS or RR fluxes.\\

Consider the NS-NS and RR potentials given by
\beq
B_2= b_2+ b^a\omega_a+(b^in_i)\wh\omega_2 \quad \text{and} \quad C_2=c_2+ c^a\omega_a+(c^in_i)\wh\omega_2,
\eeq
where $b^i$ and $c^i$ are constant real scalar fields and $b_2$ and $c_2$ are 2-forms supported in the four-dimensional extended space-time (non necessarily constants) and $\omega_a$ is the 2-form basis in $\H^2(Y_3;\mathbf{Z})$ with a constant $b^a$ and $c^a$. The corresponding field strengths for these potentials read
\begin{eqnarray}
H_3&=&dB_2=db_2+\left(k(b^in_i)+e_0\right)\beta^{0,tor},\nonumber\\
F_3&=&dC_2-C_0dB_2=dc_2+k(c^in_i)\beta^{0,tor}-C_0H_3.
\end{eqnarray}
By expanding the RR potential $C_4$ as
\beq
C_4=A_1\wedge \left[ e^i\alpha_i-m_i\beta^i +\e\a-m_0\b\right],
\eeq
the 5-form field strength reads
\beq
F_5=F_2\wedge\left[ e^i\alpha_i-m_i\beta^i+ \e\a-m_0\b\right]-A_1e^0\omega_4^{tor}.
\eeq
Comparing this expression for $F_5$ with that given in (\ref{F5D3}), we see that the last term in the right hand side does not contribute to the charge and that is why it was not considered in the calculation of the black hole charge, although it plays an important role in the low-energy effective theory we are computing.\\

Therefore, using the above expressions for $C_4$ and $F_5$, together with the complex moduli $Z^i$ and $v^i$ from equations (\ref{omega}) and (\ref{J}) it is possible to construct the multiplets for the effective theory ${\cal N}=2$ in four dimensions where the gravity multiplet consists on the graviton $g_{\mu\nu}$ and the vector field (graviphoton) $\e A_1$. The vector multiplet is given by $(e^aA_1, Z^a)$ with $a=1,\cdots , h^{(2,1)}(Y_3)$ and the scalars conforming the hyper-multiplets $(\phi, C_0, \star b_2, \star c_2, b^i, c^i, v^i)$ where $\star$ is the Hodge-dual in four dimensions.\\

Not being a Ricci-flat manifold, compactification on a half-flat manifold  \cite{Gurrieri:2002wz} leads to an effective potential induced by  the internal curvature  and given by
\beq
V_g^{HF}=-\frac{\kappa_0}{16{\cal K}}e^{2\phi}k^2n_in_jg^{ij},
\eeq
where $g^{ij}$ is the metric of the scalar moduli space. After incorporating self-duality on $F_5$, it was shown that \cite{Gurrieri:2002iw} some of the fields carry non-physical degrees of freedom. In particular it is possible to show that $m_IF_2$ can be eliminated. Notice that this is compatible with cancelation of Freed-Witten anomaly once $D3$-branes are considered as in our case. Therefore the low energy action reads:
\beq
S_{IIB}=\int -\frac{1}{2}R*\mathbf{1}+\frac{1}{2}Im~{\cal M}_{IJ}F^I\wedge*F^J+\frac{1}{2}Re~{\cal M}_{IJ}F^I\wedge F^J-h_{\mu\nu}Dq^\mu\wedge\*Dq^\nu-V_{eff}*\mathbf{1},
\eeq
where, following the notation in \cite{Gurrieri:2002iw}, $q=(\phi, a, \xi^I, \tilde\xi_I)$ with the scalar fields  given by
\begin{eqnarray}
a&=&2\star b_2+C_0\star c_2,\nonumber\\
\xi^I&=&(C_0, C_0b^i-c^i), \nonumber\\
\wt\xi_I&=&(-\star C_2-\frac{C_0}{6}{\cal K}_{ijk}b^ib^jb^k+\frac{1}{2}{\cal K}_{ijk}b^ib^jb^k, \frac{C_0}{2}{\cal K}_{ijk}b^jb^k-{\cal K}_{ijk}b^jc^k),
\end{eqnarray}
and the covariant derivatives read
\begin{eqnarray}
D\tilde\xi_I&=&d\wt\xi_I-k\e n_IA_1,\nonumber\\
Da&=&da+k\e A_1n_I\xi^I.
\end{eqnarray}
with $d\wt\xi_I=(-d\star C_2, \frac{dC_0}{2}{\cal K}_{ijk}b^jb^k)$. Then it follows that
\beq
h_{\mu\nu}Dq^\mu\wedge*Dq^\nu=d\phi\wedge*d\phi+g_{ab}dZ^a\wedge d\bar{Z}^b+{\cal L}_{mscalars} + {\cal L}_{mgauge},
\eeq
with
\begin{eqnarray}
{\cal L}_{mscalars}&=&\frac{e^{4\phi}}{4}\left[Da-\xi^I(d\wt\xi_I-kn_I\e A_1)\right]\wedge * \left[Da-\xi^I(d\wt\xi_I-kn_I\e A_1)\right],\nonumber\\
{\cal L}_{mgauge}&=&-\frac{e^{2\phi}}{2}C^{IJ}(d\wt\xi_I-kn_I\e A_1)\wedge*(d\wt\xi_J-kn_J\e A_1).
\end{eqnarray}
Notice then than the gaugings from ${\cal L}_{mscalars}$ show that the scalar $a$ is charged under the graviphoton and also that the scalars $b^i$ and $c^i$ become massive through the terms $\xi_I$. Concerning the Lagrangian term ${\cal L}_{mgauge}$, this is exactly a St\"uckelberg Lagrangian, showing that the involved photons $\e A_1$ are massive \cite{Camara:2011jg}.\\

Finally let us comment on the effective scalar potential given by
\beq
V_{eff}=V_g^{HF}-\frac{\kappa_0}{2}e^{4\phi}(kn_I\xi^I)^2+\frac{V_{BH}}{r^4},
\eeq
where $V_{BH}$ is given by expression (\ref{VBH}). The stability of such system depends essentially on parameters $n_I$ (coming from the mirror symmetric image of electric NS-fluxes in Type IIA on a CY manifold $X_3$), the curvature of the internal manifold and the contribution of the supersymmetric part of the black hole through the term $V_{BH}$. The stability of a SBH in a background threaded with fluxes was studied in \cite{Danielsson:2006jg}. For the present case treated in this note, we leave the study of this important fact for a future work.\\

\bibliography{K-QH}
\addcontentsline{toc}{section}{Bibliography}
\bibliographystyle{TitleAndArxiv} 
\end{document}